# Terrestrial Planet Formation: Constraining the Formation of Mercury

Patryk Sofia Lykawka[1] and Takashi Ito (伊藤 孝士)[2]

[1] School of Interdisciplinary Social and Human Sciences, Kindai University, Shinkamikosaka 228-3, Higashiosaka, Osaka, 577-0813, Japan; patryksan@gmail.com

[2] National Astronomical Observatory of Japan, Osawa 2-21-1, Mitaka, Tokyo, 181-8588, Japan

## ABSTRACT

The formation of the four terrestrial planets of the solar system is one of the most fundamental problems in the planetary sciences. However, the formation of Mercury remains poorly understood. We investigated terrestrial planet formation by performing 110 high-resolution N-body simulation runs using more than 100 embryos and 6000 disk planetesimals representing a primordial protoplanetary disk. To investigate the formation of Mercury, these simulations considered an inner region of the disk at 0.2–0.5 au (the Mercury region) and disks with and without mass enhancements beyond the ice line location, $a_{IL}$, in the disk, where $a_{IL}$ = 1.5, 2.25, and 3.0 au were tested. Although Venus and Earth analogs (considering both orbits and masses) successfully formed in the majority of the runs, Mercury analogs were obtained in only nine runs. Mars analogs were also similarly scarce. Our Mercury analogs concentrated at orbits with $a$ ~ 0.27–0.34 au, relatively small eccentricities/inclinations, and median mass $m$ ~ 0.2 $M_{\oplus}$. In addition, we found that our Mercury analogs acquired most of their final masses from embryos/planetesimals initially located between 0.2 and ~1–1.5 au within 10 Myr, while the remaining mass came from a wider region up to ~3 au at later times. Although the ice line was negligible in the formation of planets located in the Mercury region, it enriched all terrestrial planets with water. Indeed, Mercury analogs showed a wide range of water mass fractions at the end of terrestrial planet formation.

Key words: methods: numerical — planets and satellites: dynamical evolution and stability — planets and satellites: formation — planets and satellites: general — planets and satellites: individual: Mercury — planets and satellites: terrestrial planets

# 1. INTRODUCTION

The terrestrial planets in the inner solar system probably formed on timescales of 10–100 Myr, presumably after the four giant planets were fully formed in the system (Chambers 2007; Morbidelli et al. 2012; Raymond et al. 2014). Terrestrial planet formation can be summarized in three main stages: growth of small planetesimals, runaway growth of small embryos, and oligarchic/chaotic growth of (larger) embryos. The first two stages are the least understood; researchers have proposed many mechanisms for them, including cumulative accretion (e.g., Kokubo & Ida 2000; Goldreich et al. 2004), gravitational collapse (e.g., Morbidelli et al. 2009), and pebble accretion (Levison et al. 2015; Chambers 2016). In the last stage, the resulting system typically yields tens to hundreds of embryos and a remaining population of small planetesimals. Ultimately, embryo–embryo collisions and further accretion of planetesimals by the embryos lead to a stable system of a few final terrestrial planets after a few hundred Myr, a common result obtained by various research groups (Chambers 2001; O'Brien et al. 2006; Morishima et al. 2010; Raymond et al. 2009; Lykawka & Ito 2013; Fischer & Ciesla 2014; Izidoro et al. 2015; Brasser et al. 2016; Haghighipour & Winter 2016).

With the goal of reproducing the orbits and masses of the terrestrial planets of the solar system, previous studies explored several parameters in their initial conditions, such as the orbital architecture of Jupiter and Saturn (e.g., current orbits vs. orbits within ~15 au) and the mass distribution of the protoplanetary disk in the early inner solar system (e.g., extended disks up to ~4.5 au vs. narrow disks truncated at ~1 au). These studies found that systems in which Jupiter and Saturn started on eccentric orbits (comparable to or slightly higher than their current values) tended to produce systems resembling the system of terrestrial planets (O'Brien et al. 2006; Raymond et al. 2009; Walsh & Morbidelli 2011). However, there were still some unsolved problems, such as highly inefficient water delivery to Earth and obtained Mars-like planets that were at least a few times more massive than Mars. More recent studies found that simulations starting with narrow protoplanetary disks better address these issues, suggesting that disk properties hold the key to producing systems similar to that of the solar system terrestrial planets (Hansen 2009; Jacobson & Morbidelli 2014; Walsh & Levison 2016).

Despite the efforts of these studies, reproducing *the orbits and masses of all four terrestrial planets* remains elusive. First, the formation of Mercury has received little attention in any previous study. Indeed, in most cases, it is completely omitted. This is not surprising. First, the formation of Mercury may require more complicated modeling with the protoplanetary disk featuring an inner region component (e.g., Jacobson & Morbidelli 2014). Chambers (2001) and O'Brien et al. (2006) considered such a region at $a = 0.3$–$0.7$ au in simulations using disks extended out to 2 and 4 au, respectively. Chambers (2001) found a few planets with orbits similar to that of Mercury ($a \sim 0.5$ au) and masses in the range 0.2–0.5 $M_\oplus$. By contrast, in the study by O'Brien et al. (2006), the initial masses of embryos were too massive to address the formation of Mercury. Hansen (2009) obtained a few planets at $a < 0.65$ au (defined as the Mercury region in that reference) using a narrow disk set at $a = 0.7$–$1.0$ au without inner components. The masses of those planets varied widely at 0.005~0.9 $M_\oplus$, but four of them were Mercury analogs with $a < 0.5$ au and masses within 0.05–0.2 $M_\oplus$. Finally, Izidoro et al. (2015) obtained a few planets at $a \sim 0.5$ au with masses 0.05–0.4 $M_\oplus$ using a disk starting at 0.7 au with a very steep mass surface density. Second, irrespective of setting the inner edge of the protoplanetary disk near Mercury's orbit ($a = 0.39$ au) or beyond, to obtain accurate results at $a \sim 0.4$ au, we believe that very small time steps are needed when performing simulations (i.e., <1/20 Mercury's orbital period ~ 4.5 days). To avoid simulations that were too computationally expensive, previous results typically used larger time steps, ~6–8 days. For this reason, those studies did not yield accurate results to investigate Mercury's local formation at $a \sim 0.4$ au or to address the hypothesis that the planet was scattered from outer regions to its current orbit. To our knowledge, the only study that met the above condition was that of Hansen (2009), in which the time step used was 4 days. Third, there is insufficient discussion about the likelihood of obtaining Mars analogs (with ranges of semimajor axis and mass similar to those of Mars) *and* Venus/Earth analogs *in the same systems* (Raymond et al. 2009; Jacobson & Walsh 2015; Brasser et al. 2016). Furthermore, additional open questions arise if we consider other constraints that must be satisfied in the inner solar system (see discussion in Lykawka & Ito 2013; Jacobson & Morbidelli 2014; Izidoro et al. 2015). For these reasons, despite the important advances in terrestrial planet formation science that have been made in the last few years, further studies are crucial for improved understanding of the problems mentioned above. In particular, we believe that more systematic studies exploring the architectures of the giant planets and the various properties of the protoplanetary disk are warranted.

In this study, we perform simulations of terrestrial planet formation to clarify the formation of Mercury and the influence of the ice line in the protoplanetary disk that formed the four terrestrial planets. Although the focus of this work is on Mercury and the inner terrestrial planet region (i.e., <1.5 au), we intend to consider the formation of Mars analogs and disks with different properties in upcoming work. Our main goals are the following:

*Mercury*. We aim to understand better how the planet could have formed in the context of formation of the other terrestrial planets. In particular, Mercury could have formed as a relatively large embryo that was scattered to its current orbit from a more distant region (external origin). Alternatively, an embryo that formed near the current orbit of Mercury may have collided with neighboring embryos and planetesimals to form the planet (local origin). In addition, we aim to provide new constraints on the formation of Mercury, such as the main source region of Mercury-like planets, the delivery of water to the planet, and others. Finally, the recent discovery of water ice and other volatile deposits on Mercury indicates that the origin of these substances is an important issue strongly connected to the origin of the planet (Benz et al. 2007; Peplowski et al. 2011; Eke et al. 2017).

*Ice line*. In this paper, we consider the ice line to be the distance from the Sun beyond which water solidifies in space (water ice). In the current solar system, the location of the ice line is at approximately 2.7 au, but it probably varied within the range of 1–3 au during the early system (Lunine 2006; Dodson-Robinson et al. 2009). We aim to understand how the enhancement of mass in the protoplanetary disk beyond the ice line may have influenced the formation of Mercury and the other planets in the inner solar system. In particular, it may offer new hints about the delivery of water to Mercury-like planets. Indeed, the origin of water on all of the terrestrial planets may be connected to the nature of the ice line (e.g., Sato et al. 2016).

To address the main goals mentioned above and unveil potential new insights/constraints, our simulations contain a few new features and improvements compared with previous studies. First, we explored the parameter space with at least 10 runs per specific initial condition for better statistics. Also, all simulations used 100+ embryos and 6,000 planetesimals in each run, thus satisfying the condition that the inner solar system protoplanetary disk should be modeled with sufficient resolution (>2,000–3,000 planetesimals) for accurate representation of the system dynamics (Morishima et al. 2008). We also verified the importance of the disk mass distribution between embryos and planetesimals by testing three ratios of total disk mass in embryos to mass in planetesimals in all simulations: embryo-dominated, planetesimal-dominated, or equal contributions. Finally, for a quick evaluation of the system outcome, we define the Mercury, Venus, Earth, and Mars regions as those located at $a = 0.2$–0.5 au (Mercury), 0.5–1.2 au (Venus–Earth), and 1.2–2.0 au (Mars). Any object with mass > 0.05 $M_\oplus$ found within its region after 1 Gyr of orbital evolution is defined as planet-like, whereas a planet analog further requires mass within a factor of two of the mass of the real planet (we relax this condition for Mercury; see Section 3.1). Finally, systems containing both analogs of Mercury and Venus are deemed successful for the purposes of this work (see Section 3.3 for details).

## 2. METHODS

The starting conditions of the simulations performed in this work are assumed to represent the solar system when the disk gas mass became negligibly small, as compared to the mass in solids, corresponding to roughly the time of a few Myr after the birth of the system (Haisch et al. 2001; Gorti et al. 2016). In particular, we consider a primordial system consisting of Jupiter and Saturn and a disk consisting of hundreds of embryos and thousands of remnant planetesimals

Because Jupiter and Saturn interact with the outer region of the primordial disk, they tend to migrate slightly inward from their initial locations. Also, these interactions usually lead to modest planetary eccentricity damping during the first tens of Myr (e.g., Lykawka & Ito 2013). Thus, for Jupiter and Saturn to acquire orbits similar to their current orbits at the end of terrestrial planet formation, both planets must start with orbits slightly farther away and more eccentric than their current orbits[1]. For instance, the giant planets could have acquired such orbits after suffering dynamical instabilities in the early solar system (Ford & Chiang 2007; Thommes et al. 2008; Nesvorny & Morbidelli 2012). To meet the above condition, we placed Jupiter (J) and Saturn (S) at $a_{J0} = 5.25$ au and $a_{S0} = 9.58$ au, with $e_{J0} = e_{S0} \sim 0.09$, i$J0 \sim 1.5$ deg, and $i_{S0} \sim 2.5$ deg (i.e., similar to the EEJS configuration in Raymond et al. 2009). This choice of orbital parameters is motivated by keeping the model as simple as possible, thus avoiding uncertainties involving the behavior of the giant planets before and during their migration. Another reason for this choice is that more eccentric orbits of Jupiter and Saturn allow a faster disk dynamical evolution driven by secular resonances (O'Brien et al. 2007; Raymond et al. 2009; Haghighipour & Winter 2016), thus reducing the high computational cost of these simulations. In addition, these secular resonances could facilitate the clearing of the region beyond Mars in the primordial disk of massive embryos or of additional planets that would be in conflict with observational constraints (O'Brien et al. 2007). The influence of other giant planet orbital architectures will be presented in upcoming work.

---

[1] This may not be the case if Jupiter and Saturn acquired their current orbits after terrestrial planet formation was mostly complete (e.g., after 200 Myr or so) or if the disk was strongly depleted at the start of terrestrial planet formation (e.g., disks with mass concentrations within 1–1.5 au).

Our disk of planetesimals and embryos was initially set within 0.2–3.8 au, with a total mass of ~7 $M_\oplus$ for standard disks and ~9.4–15.3 $M_\oplus$ for disks with ice lines. Similar to the model described in O'Brien et al. (2006), the distribution of mass in the disk resulted in a surface density that obeyed a linear increase from 0.2 to 0.5 au (the Mercury region) and a decay power law with exponent −1.5 from 0.5 to 3.8 au. To avoid overpopulating the disk with too many small embryos near the disk inner edge at 0.2 au, the disk mass in the $a = 0.2$–0.25 au region was represented solely by planetesimals. By extending the disk to the Mercury region, we can better understand the role of the latter in the formation of Mercury and address its influence beyond 0.5 au. To address the effects of disk mass enhancements beyond the ice line, we also considered disks with ice lines located at $a_{IL} = 1.5$, 2.25, and 3.0 au, beyond which the surface density was increased by a factor of 3 (Garaud & Lin 2007; Dodson-Robinson et al. 2009; Min et al. 2011). The disk outer edge (3.8 au) was determined by the condition $a_{J0}(1 - e_{J0}) - 3R_{H,J0}$, where $R_{H,J0}$ is the Hill radius of Jupiter at the start of the simulations. Test simulations show that disk objects located within $3R_H$ of a planet are very quickly removed from the system, so we assume that these objects were already gone at the start of these simulations.

We tested three distributions of disk mass represented by the ratios of total mass in embryos to that in planetesimals of $r = 0.5$ (planetesimal-dominated), 1 (equal contributions), and 4 (embryo-dominated). All disks started with ~0.45 $M_\oplus$ in the Mercury region at $a = 0.2$–0.5 au and ~2 $M_\oplus$ in the Venus–Earth region at $a = 0.5$–1.2 au. The disk total mass was distributed between 6,000 planetesimals (with ~5,000 placed beyond 0.5 au) and 132–216 embryos, where the latter varied depending on the setup (Table 1). The large number of planetesimals also allowed dynamical friction to work even for the smaller embryos. The mass distribution of embryos followed the typical result of oligarchic theory (Leinhardt & Richardson 2005; Kokubo et al. 2006). Embryos started in dynamically cold orbits with $e_0 < 0.01$ and $i_0 \sim e/2$ (~0.3 deg), while planetesimals started with $e_0 < 0.03$ and $i_0 \sim e/2$ (~1 deg). The initial eccentricities and inclinations were chosen randomly within these ranges. The initial semimajor axis of the embryos was determined according to $a_{n+1} = a_n + bR_{mH}$, where b ~ 5.5, $R_{mH} = [(a_{n+1} + a_n)/2] \cdot [(m_{n+1} + m_n)/3M]^{\wedge}(1/3)$ (e.g., Chambers & Wetherill 1998), $m$ is the mass of any two adjacent embryos, $M$ is the mass of the Sun, and $R_{mH}$ is the mutual Hill radius of the two embryos. To avoid using too many small embryos, according to the law above, the embryos started at $a = 0.25$ au with minimum masses of ~$8 \cdot 10^{-5}$ to $3 \cdot 10^{-4}$ $M_\oplus$, depending on the $r$ considered in the model. Similar initial conditions were used in several previous terrestrial planet formation models (Chambers 2001; O'Brien et al. 2006; Raymond et al. 2009; Morbidelli et al. 2012; Raymond et al. 2014; Izidoro et al. 2015; Brasser et al. 2016). The bulk density of planetesimals and embryos was 3 g cm$^{-3}$. Typical initial conditions for the planetesimals and embryos are illustrated in Fig. 1.

To address the amount of water accreted in the planets formed, we used a simple water model based on Morbidelli et al. (2012). In this model, the water mass fraction (WMF) of embryos/planetesimals was defined according to their initial location in the protoplanetary disk as follows: $10^{-5}$ at $a \leq 1.5$ au, $10^{-4}$ at $1.5 \leq a < 2$ au, $10^{-3}$ at $2 \leq a < 2.5$ au, $5 \cdot 10^{-2}$ at $2.5 \leq a < 3$ au, and $10^{-1}$ at $a \geq 3$ au. For disks with ice lines, the region with WMF = $10^{-1}$ was set at $a > a_{IL}$.

Eleven simulation scenarios, with 10 runs of each given scenario, were considered to investigate the formation of Mercury-like planets and to understand the influence of the $a_{IL}$ and $r$ parameters described above. All calculations were performed using a modified version of the MERCURY integrator (Chambers 1999; Hahn & Malhotra 2005). Jupiter, Saturn, and embryos are modeled as massive bodies that fully perturb each other, including collisions. Planetesimals are also considered to be massive and to interact with planets and embryos, but they do not feel perturbations from each other. Collisions were considered to be perfectly inelastic. All simulations were evolved until 1 Gyr. The basic time-step used was 1/225 yr (~1.6 days). Bodies that acquired heliocentric distances smaller than 0.08 au or larger than 40 au were eliminated from the simulations. A summary of the simulations is given in Table 1.

## 3. MAIN RESULTS AND DISCUSSION

As in previous investigations that used similar modeling, we obtained a number of systems with typically 3–5 terrestrial planets in each system. Figure 2 summarizes the combined results of all main simulations, and Fig. 3 shows the same results in terms of individual obtained systems. Overall, Venus and Earth analog planets are commonplace and show both similar masses and orbits, whereas Mercury and Mars analogs are scarce. These results and trends are in agreement with those of past studies (e.g., Morbidelli et al. 2012; Raymond et al. 2014). Furthermore, systems with higher $r$ and smaller $a_{IL}$ tended to form fewer planets and planets on dynamically more excited orbits (notably, higher eccentricities) compared with other systems. The reason is twofold. Systems with higher $r$ had more disk mass distributed among embryos, resulting in initially larger embryos. Conversely, with less mass in planetesimals in those systems, larger embryos tended to perturb each other more strongly while "feeling" less the cooling effects of dynamical friction by the planetesimals. Finally, systems with smaller $a_{IL}$ allowed the existence of quite large icy embryos beyond the location of the ice line (Fig. 1); thus, these objects perturbed each other and the system as a whole. Such stronger perturbations led the embryos to experience more giant impacts, collisions with Jupiter or the Sun, and ejections from the system after acquiring unstable orbits. Although disks with an ice line experienced much stronger dynamical evolution with embryo–embryo scattering and giant collisions beyond ~1.5–2 au, these effects were not strong enough to deplete that region in a way that would efficiently produce Mars analogs.

### 3.1 Mercury-like planets

We identified 52 planets in the Mercury region (Mercury-like planets) from the combined results obtained in this work, out of 110 runs. As shown in Table 2, in general, these planets are several times as massive as Mercury. These planets also typically concentrate closer to the Sun than Mercury, at $a \sim 0.33$–$0.39$ au, and tend to have smaller masses for orbits with smaller semimajor axes. Lastly, we noticed that the larger the value of $r$, the smaller the number of Mercury-like planets obtained in the final systems (Fig. 2 and Table 2). These results suggest that considering a Mercury region located slightly beyond the previously used value of $a = 0.2$–$0.5$ au with $r \leq 1$ would increase the chances of reproducing the orbit of Mercury.

Our Mercury-like planets typically started as small embryos (with an initial mass of only a few % that of the planet's final mass) but later acquired 90% of their mass through the accretion of neighboring embryos and planetesimals located between 0.2 and 0.8–1.6 (mostly ~1.3) au *for all scenarios tested* (Table 2). Therefore, the disk mass/structure beyond ~1.5 au should play a minor role in determining the orbit and mass of Mercury. These results also suggest that disks with smaller mass distributions beyond 1.5 au (e.g., truncated or with steep mass surface density) will play a negligible role in this regard.

As expected, the contribution of embryos (planetesimals) to the final planets increased (decreased) with $r$, while for systems with equal initial distribution of embryos/planetesimals ($r = 1$), the contribution of embryos (planetesimals) was found to be at the 30–40 (60–70)% level. Finally, of the ~0.45 $M_\oplus$ that was initially available in the Mercury region, approximately 40–60% of that mass was typically incorporated into the Mercury-like planets that formed. This contribution alone exceeds the mass of Mercury by 3–5 times, so a strongly or totally disk-mass depleted Mercury region is required to form a small-mass Mercury. Still, the mass incorporated into those planets from the region of 0.5–1.5 au was similarly a few times the mass of Mercury. This suggests that although a disk starting at 0.5 au without including the Mercury region may satisfy the formation of Venus and Earth analogs, the same disk would not be able to produce Mercury analogs consistently, because the latter would be too massive[2]. Therefore, investigations of disks with Mercury regions set with different inner/outer edges and total masses are warranted.

### 3.2 Mercury analogs

By setting the mass requirement for Mercury analog planets to a maximum of 0.25 $M_\oplus$, we can identify nine analogs from the 52 Mercury-like planets (see Table 3 for a summary). As detailed below, the general trends discussed for the planets obtained above are also seen for the Mercury analogs. No strong preference of disk scenario was found: standard disks, disks with $a_{IL} = 2.25$ au, and disks with $a_{IL} = 3.0$ au yielded 2, 4, and 3 Mercury analogs, respectively (Table 3 and Fig. 4). Disks with an ice line located at 1.5 au did not produce Mercury analogs, but rather produced only Mercury-like planets with at least 0.35 $M_\oplus$. Thus, this particular scenario is not considered in the discussion below. Also, Figs. 4 and 5 illustrate that the growth timescales and feeding zones of the Mercury analogs showed no obvious dependence on disk scenario used.

Our Mercury analogs concentrated at orbits with $a \sim 0.27$–0.34 au, relatively small eccentricities/inclinations, and median mass $m \sim 0.2$ $M_\oplus$. Eight Mercury analogs originated at 0.31–0.43 au and acquired their bulk masses in tens of Myr in systems with $r = 0.5$ and $r = 1$ (Fig. 4). All of the Mercury analogs except for one are composed mostly of aggregated planetesimals, and ~50–70% of their final masses originated in the Mercury region at 0.2–0.5 au. Similarly to Mercury-like planets, 90% of the accreted final mass of our Mercury analogs originated from objects initially located between 0.2 and 0.9–1.4 au (median 0.2 and 1.1 au), with a notable peak at 0.2–0.6 au. The remaining minor contribution to the final planets' masses came from embryos/planetesimals initially located between approximately ~1–1.5 au and 3 au (Fig. 5). Thus, similar to the findings presented in Section 3.1, the formation of our Mercury analogs implies that the region beyond ~1–1.5 au plays a minor role in reproducing Mercury's orbit and mass.

---

[2] However, note that simulations including fragmentation may allow the formation of less massive Mercury-like planets (Chambers 2013). Thus, the values presented here are upper limits for the model conditions used in this work.

As shown in detail in Fig. 4, the obtained Mercury analogs acquired masses comparable to that of Mercury (0.055 $M_\oplus$) in less than 10 Myr and later further acquired ~0.1–0.2 $M_\oplus$ of mass during the next ~100 Myr. These planets exhibited spikes in their orbital evolution due to giant collisions and embryo–embryo gravitational scattering within ~10–30 Myr. Nevertheless, in general, their eccentricities and inclinations remained below 0.3 and 10 deg, respectively. Lastly, the obtained Mercury analogs also migrated slightly inwards from their initial location in the disk.

Overall, the aforementioned results suggest that forming a small-mass Mercury at $a \sim 0.4$ au would require a disk with $r \leq 1$ containing a less massive Mercury region placed at a larger distance from the Sun during the early solar system. In particular, extending the region's outer edge beyond 0.5 au would reduce the mass of the feeding zone of Mercury-like planets. Also, the orbital decay of the Mercury analogs would favor a Mercury region with an inner edge at least ~0.1 au greater than that used here. Therefore, for example, a Mercury region set with inner edge $a \geq 0.3$ au and outer edge $a \geq 0.6$ au would probably lead to improved results. Other possible scenarios include disks containing Mercury regions with mass depletion.

### 3.2.1 Water delivery and clues about Mercury's origin

Although the majority of the mass accreted into the Mercury analogs came from the region within ~1–1.5 au, we found a non-negligible contribution of water due to accretion of water-rich embryos/planetesimals initially located beyond 1.5 au. As detailed in Section 2, those objects had WMFs larger than $10^{-5}$, which was the initial value of all our Mercury-like planets. Tables 2 and 3 indicate that the Mercury-like and analog planets acquired a wide range of WMFs by the end of the simulations, generally higher than $10^{-5}$. This variation of WMF resulted from the stochastic nature of Mercury-like and analog planets experiencing sporadic collisions with water-rich objects, which were in particular dominated by a few collisions of objects initially located within 2–3 au. Note that because Mercury's WMF has not yet been estimated by observations, a quantitative comparison with these results is not possible. Also, the water model used in this work is not unique; thus, more work is needed to address in greater detail the water contents acquired by Mercury during its formation. Nevertheless, these results indicate that the delivery of water to Mercury is a natural outcome during terrestrial planet formation.

Considering the origin of Mercury (local vs. external), as discussed in Section 1, the overall results above suggest that Mercury probably did not form locally based on the initial conditions used in this work. Even our best Mercury analogs were found to be somewhat too massive. Nevertheless, clues on the planet's origin can be achieved from those results. First, we did not observe a Mercury analog originating beyond ~0.6 au, which suggests that an external origin would be less likely for Mercury. Next, assuming the existence of a compositional gradient in the primordial inner solar system, our Mercury analogs accreted most of their mass from the iron-rich 0.2–0.6 au region and, to a lesser extent, from the silicate-rich 0.6–1.5 au region (see Fig. 5). It is worth noting that the accretion of materials from the region beyond ~1.5 au occurred mostly at late times (>10 Myr). Thus, this Mercurian late veneer could explain the presence of volatile-rich materials in Mercury's crust and surface, as was found by the MESSENGER mission (e.g., Peplowski et al. 2011). The formation of Mercury in an iron-rich local region along with a late veneer enriching the planet with water and other volatiles represents a scenario worth investigating in the future.

### 3.3 Planetary systems analogous to the inner solar system?

Although our models were not designed to reproduce the main features of the four terrestrial planets, we obtained a few individual planetary systems that seem to resemble the inner solar system in terms of orbits and masses. If the existence of a Mercury analog ($m < 0.25$ $M_\oplus$) is considered a measure of success, models Sr1, IL2r05, IL2r1, IL3r05, and IL3r4 would become representative scenarios for Mercury formation (see Fig. 3 and Table 3). By further requiring the existence of a Venus analog in the same system with $a > 0.5$ au and WMF $> 5 \cdot 10^{-5}$ (footnote [3]), the best model systems would be IL2r05 (systems 4 and 6), IL2r1 (system 2), IL3r05 (system 9), and IL3r4 (system 6). However, a caveat is that both our Mercury and Venus analogs obtained in the same systems orbit closer to the Sun than the real planets (Tables 3 and 4). Furthermore, the dynamical chaotic nature of Mercury suggests that not all of our Mercury analogs would be stable over the age of the solar system (Ito & Tanikawa 2002; Laskar & Gastineau 2009); thus, long-term investigation of obtained planetary systems is also important.

Based on the findings of this work, the exploration of additional similar disk parameters could lead to further improved results. In particular, important improvements include the following: 1) simultaneous formation of lower-mass Mercury and Mars-like planets; 2) better orbital concentration of Venus- and Earth-like planets at $a = 0.7$–1 au in systems with Mercury analogs; and 3) non-formation or removal of unwanted overly massive bodies beyond 2 au. Concerning point 1, disks with inner edges at 0.3 au or beyond and narrower disks (e.g., truncated at 1–2 au or with steeper mass surface densities) could yield better results. With regard to point 2, disks with a peak disk mass set at 0.7 au or beyond (instead of 0.5 au, as in this work) are likely to solve this problem (see also the discussion in Brasser et al. 2016). Finally, point 3 requires exploring narrower disks. Furthermore, alternative orbital architectures for the giant planets are particularly likely to affect the results pertaining to points 1 and 3.

In addition to the orbits and masses of the four terrestrial planets, forming planetary systems analogous to the inner solar system requires satisfying other constraints, such as the amount of water delivered to the planets, the timing of the giant collision that created the Earth–Moon system, and the late delivery of materials to Earth (after Moon formation), among others (Lykawka & Ito 2013; Jacobson & Morbidelli 2014; Izidoro et al. 2015). A more detailed analysis considering these constraints and using more rigorous system indicators, such as the radial mass concentration and the angular momentum deficit, will be conducted in future work.

---

[3] It has been suggested that after Venus formation, the planet probably had a WMF at least comparable to Earth's current ocean mass (~2.5·10$^{-4}$) (e.g., Morbidelli et al. 2000; Kulikov et al. 2006; Hashimoto et al. 2008). However, the uncertainties in those estimates are still large. Here, we consider a Venusian WMF $> 5 \cdot 10^{-5}$ to be an acceptable approximation.

## 4. CONCLUSIONS

We performed high-resolution simulations of formation of the terrestrial planets during the early solar system using primordial protoplanetary disks containing 100+ embryos and 6,000 planetesimals. In particular, we investigated the formation of planets with orbits similar to that of Mercury by considering an inner region of the disk at 0.2–0.5 au (Mercury region). We also considered disk mass enhancements caused by the ice line at distinct locations in the disk.

We obtained a number of planetary systems with 3–5 terrestrial planets located mostly within ~2 au. Our combined results formed 52 Mercury-like planets, of which 9 acquired masses within the 0.13–0.24 $M_\oplus$ range (Mercury analogs). When analyzing the orbits and masses of Mercury- and Venus-like planets as a whole, systems with 33 (67)% (i.e., $r = 0.5$) or 50 (50)% (i.e., $r = 1$) of disk mass initially distributed as embryos (planetesimals) yielded the best results when compared to the real planets.

The Mercury analogs obtained in our best terrestrial planet systems concentrated at orbits with $a \sim 0.27$–0.34 au, relatively small eccentricities/inclinations, and median masses $m \sim 0.2$ $M_\oplus$. In particular, these planets started as small embryos with only a few tenths of Mercury's mass located within 0.2–0.6 au, but later acquired their final masses from embryos/planetesimals initially located between 0.2 and ~1–1.5 au after several Myr of dynamical evolution. For this reason, the ice line played a negligible role in forming these planets. Finally, our results showed that both disks with and without an ice line were able to produce Mercury analogs with water enrichments. Thus, water delivery to Mercury is a natural outcome of terrestrial planet formation. Finally, except for models that included an ice line at 1.5 au, which did not produce Mercury analogs, we found that any of the other three scenarios tested (disks without an ice line or disks with an ice line located at 2.25 or 3.0 au) would have the potential to produce Mercury and Venus analogs.

The formation of Mercury in the context of simultaneous formation of the other terrestrial planets remains a challenging task. The best prospects for reproducing the orbit and mass of Mercury seem to require a protoplanetary disk with the following features: 1) inner edge $a \geq 0.3$ au; 2) outer edge $a \geq 0.6$ au; 3) a Mercury region less massive than ~0.45 $M_\oplus$; and 4) $r \leq 1$. Also, although the disk mass/structure beyond ~1–1.5 au would play a negligible role in forming Mercury, the same region has the potential to source Mercury with water and possibly other volatiles during a late veneer period of accretion. We will present the results of investigations considering disks with properties different from those used in this work, other giant planet orbital architectures, etc., in upcoming work.

## ACKNOWLEDGMENTS

We thank the referee for a number of helpful comments, which allowed us to improve this work. We also appreciate comments about modeling the Mercury region provided by J. C. and D. O. during the preparation of this work. All simulations presented in this work were performed using the general-purpose PC cluster at the Center for Computational Astrophysics (CfCA) in the National Astronomical Observatory of Japan (NAOJ). We are thankful for the generous time allocated to run the simulations. TI acknowledges research funding from JSPS Kakenhi Grants (JP25400458/2013-2016, JP16K05546/2016-2018) and the JSPS Bilateral Open Partnership Joint Research Project (2014–2015).

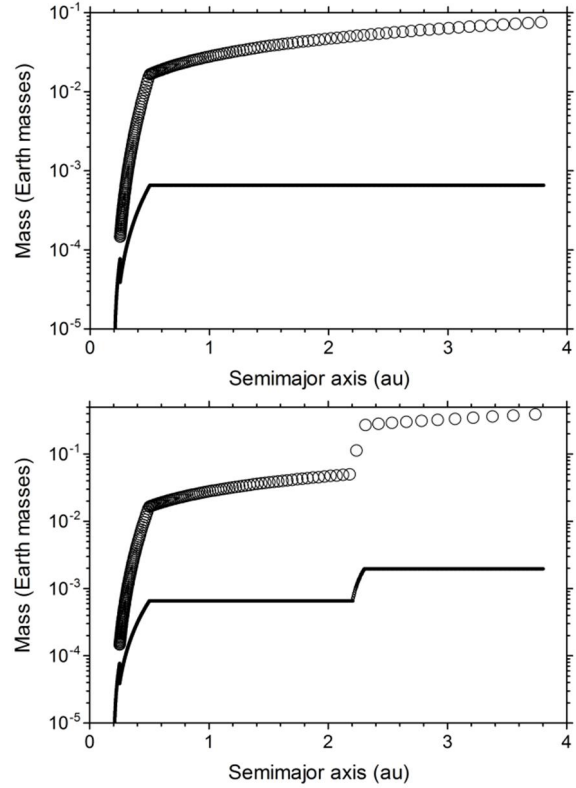

**Figure 1**. Representative initial conditions of embryos (large symbols) and planetesimals (small symbols) used in the simulations. The mass distribution in embryos/planetesimals is such that the surface density of the primordial disk increases linearly from 0.2 to 0.5 au (Mercury region) and decreases following a power law with exponent −1.5 beyond 0.5 au. Different from standard disks (top panel), disks with an ice line show a mass increase beyond the ice line location (bottom panel). The disk mass in the 0.2–0.25 au region was represented solely by planetesimals. See Section 2 (Methods) and Table 1 for more details.

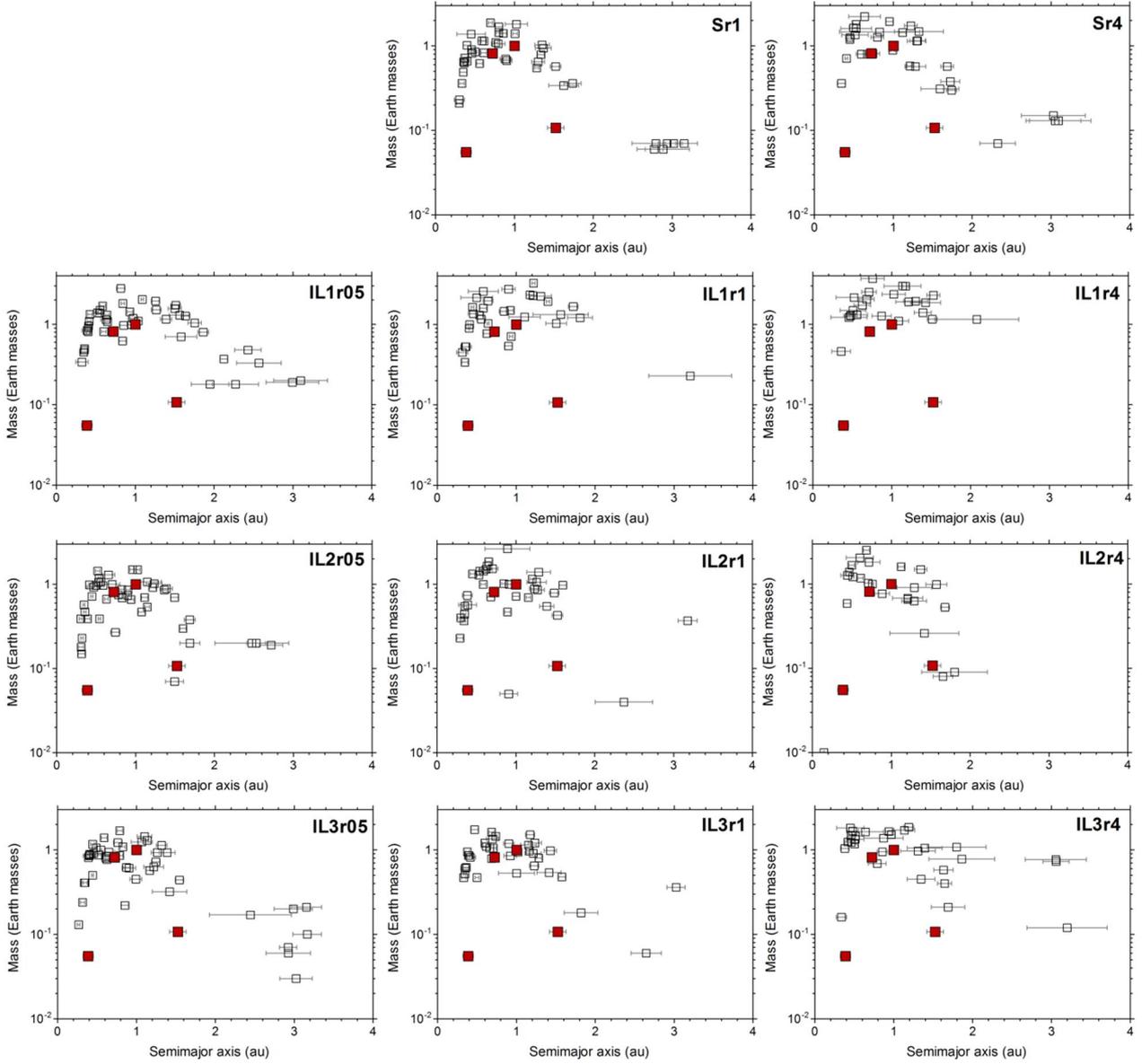

**Figure 2**. Planetary systems obtained after 1 Gyr of orbital evolution for all simulation runs performed in this work (see Table 1 for details). Red squares represent the four solar system terrestrial planets. Error bars represent the range of heliocentric distance based on the object's perihelion and aphelion.

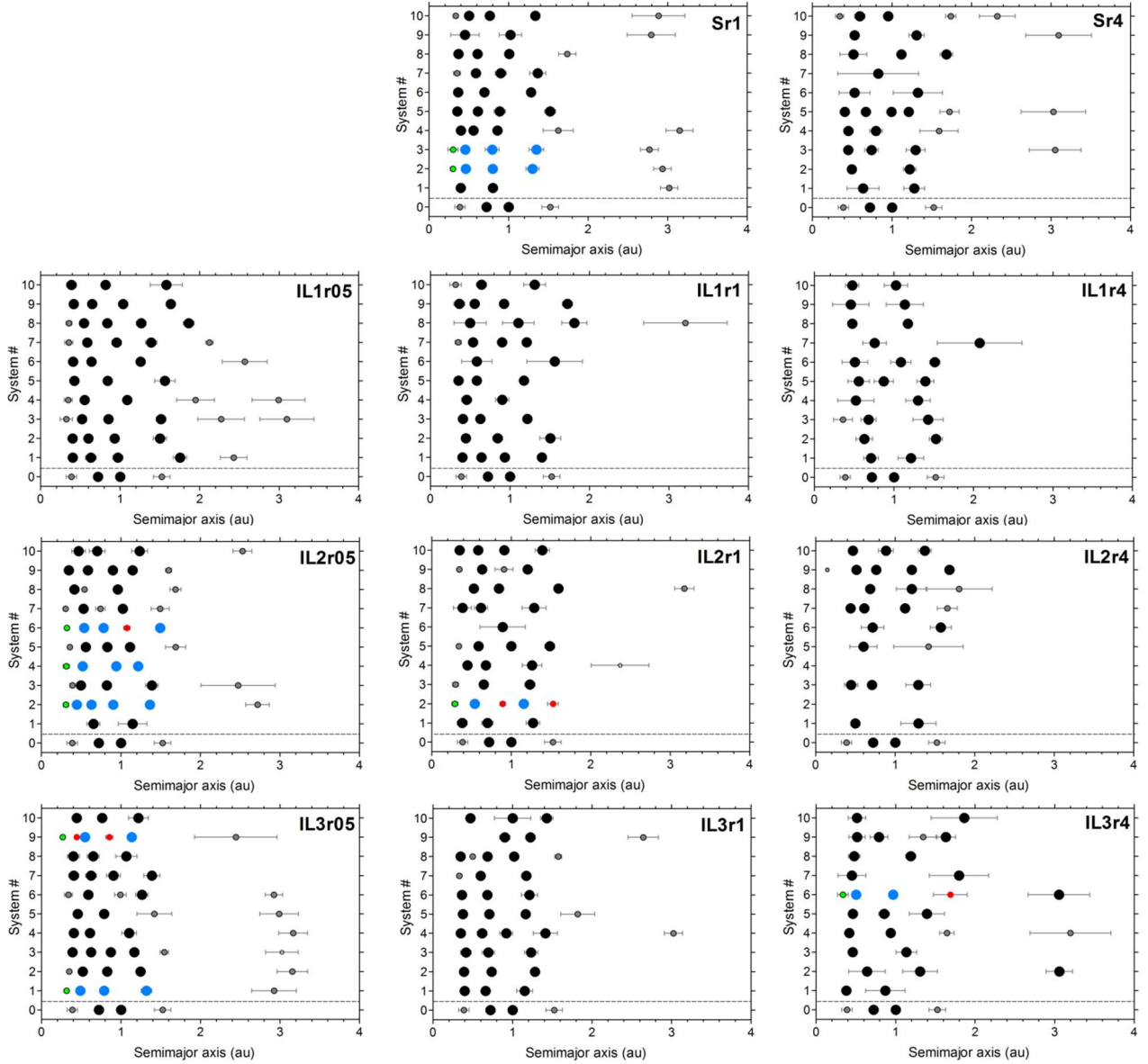

**Figure 3**. Comparison of individual planetary systems (#1–10) obtained after 1 Gyr of orbital evolution for all simulation runs performed in this work (see Table 1 for details) to the solar system planets (shown as system #0 at the bottom of each panel). More massive Venus/Earth-like planets (>0.5 M$_\oplus$) are represented by large filled circles, whereas less massive Mercury- or Mars-like planets (0.05–0.5 M$_\oplus$) are shown with gray circles. A large embryo (~0.01 M$_\oplus$) also survived in system #9 of IL2r4. Error bars represent the range of heliocentric distance based on the object's perihelion and aphelion. Finally, systems with Mercury analogs (green) are displayed in color (see Section 3.2 for details).

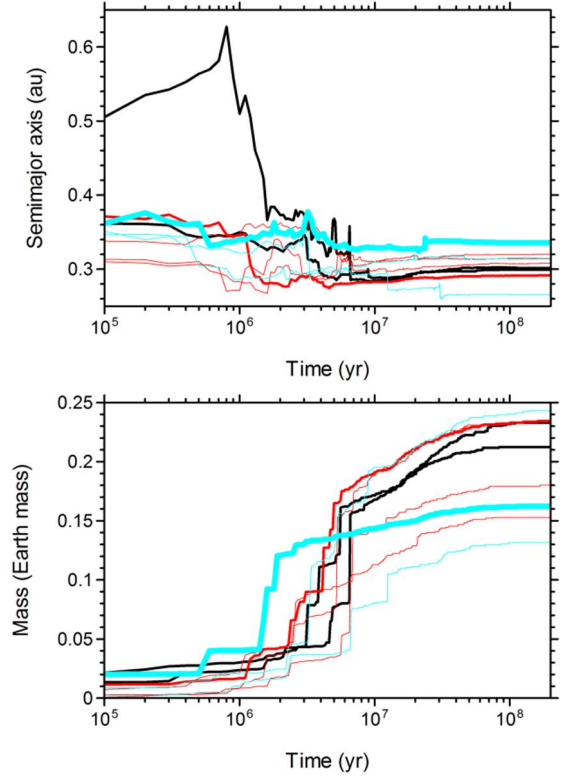

**Figure 4**. Orbital (top panel) and accretional (bottom panel) evolutions of nine Mercury analog planets obtained in each of nine individual simulation runs. The planets are indicated with different colors according to the scenario employed: standard disk without ice line (black), disk with ice line at 2.25 au (red), and disk with ice line at 3.0 au (cyan). Models with $r = 0.5$, 1, and 4 are indicated by thin, normal, and thick lines, respectively. See Table 3 and Section 3.2 for details.

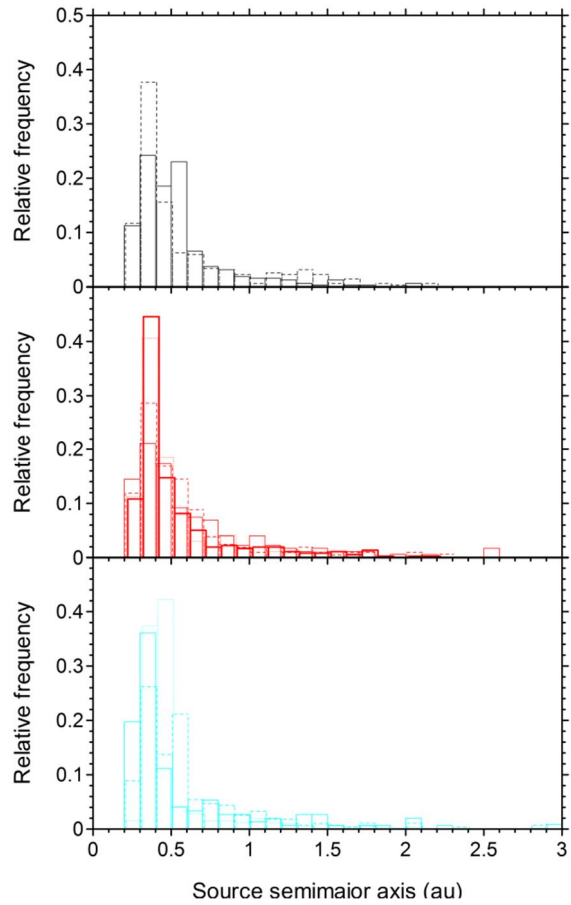

**Figure 5.** Feeding zones of the nine Mercury analog planets obtained in each of nine individual simulation runs. The planets are indicated with different colors according to the scenario employed: standard disk without ice line (black) (top panel), disk with ice line at 2.25 au (red) (middle panel), and disk with ice line at 3.0 au (cyan) (bottom panel). The bars in each panel are slightly shifted to the right for better visualization. See Table 3 and Section 3.2 for details.

**Table 1**. Main simulations

| Simulation ID | Number of embryos | $Md_{MR}$ (M$_\oplus$) | $Md_{IL}$ (M$_\oplus$) | Total disk mass (M$_\oplus$) | $a_{IL}$ (au) | $r$ | Number of runs |
|---|---|---|---|---|---|---|---|
| Sr1 | 180 | 0.46 | - | 7.03 | - | 1 | 10 |
| Sr4 | 143 | 0.46 | - | 7.01 | - | 4 | 10 |
| IL1r05 | 202 | 0.45 | 11.70 | 14.79 | 1.5 | 0.5 | 10 |
| IL1r1 | 165 | 0.46 | 11.66 | 14.74 | 1.5 | 1 | 10 |
| IL1r4 | 132 | 0.46 | 12.18 | 15.27 | 1.5 | 4 | 10 |
| IL2r05 | 210 | 0.45 | 7.19 | 11.75 | 2.25 | 0.5 | 10 |
| IL2r1 | 172 | 0.46 | 7.31 | 11.87 | 2.25 | 1 | 10 |
| IL2r4 | 137 | 0.46 | 7.40 | 11.91 | 2.25 | 4 | 10 |
| IL3r05 | 216 | 0.45 | 3.62 | 9.41 | 3.0 | 0.5 | 10 |
| IL3r1 | 177 | 0.46 | 3.80 | 9.59 | 3.0 | 1 | 10 |
| IL3r4 | 141 | 0.46 | 4.03 | 9.82 | 3.0 | 4 | 10 |

We placed embryos and planetesimals in a disk at $a$ = 0.2–3.8 au, where the disk surface density increased linearly from 0.2 to 0.5 au (the Mercury region, MR) and decayed following a power law with exponent −1.5 beyond 0.5 au. Jupiter and Saturn started at $a_{J0}$ = 5.25 au, $a_{S0}$ = 9.58 au, $e_{J0} = e_{S0}$ ∼ 0.09, $i_{J0}$ ∼ 1.5 deg, and $i_{S0}$ ∼ 2.5 deg in all simulations. Each run contained 6,000 planetesimals in the disk. All runs were followed until the total time span reached 1 Gyr. $r$ represents the ratio of total disk mass in embryos to total disk mass in planetesimals, $a_{IL}$ is the semimajor axis representing the ice line location, $Md_{IL}$ is the disk mass beyond the ice line location, and $Md_{MR}$ is the disk mass at the Mercury region located within 0.2–0.5 au.

**Table 2**. Mercury-like planets

| Sim. ID | NP | NA | $m$ ($M_\oplus$) | $a$ (au) | $e$ | $i$ (deg) | nGI | $f_0$ | $f_{Emb}$ | $f_{Obj}$ | WMF | $a_{0\_H2O}$ (au) | $a_{0\_mass}$ (au) | $f_{MR}$ |
|---|---|---|---|---|---|---|---|---|---|---|---|---|---|---|
| Sr1 | 7 | (2) | 0.49 | 0.351 | 0.101 | 4.1 | 5 | 0.03 | 0.32 | 0.65 | $8.3\cdot10^{-5}$ | 2.0-2.6 | 0.2-1.0 | 0.56 |
| Sr4 | 2 | - | 0.54 | 0.375 | 0.095 | 5.6 | 4.5 | 0.08 | 0.70 | 0.22 | $1.4\cdot10^{-5}$ | 1.5-2.4 | 0.3-1.2 | 0.40 |
| IL1r05 | 7 | - | 0.50 | 0.356 | 0.104 | 3.5 | 6 | 0.01 | 0.20 | 0.79 | $1.1\cdot10^{-2}$ | 1.5-2.4 | 0.3-1.6 | 0.40 |
| IL1r1 | 5 | - | 0.53 | 0.353 | 0.144 | 2.3 | 6 | 0.02 | 0.32 | 0.66 | $8.9\cdot10^{-3}$ | 1.5-2.7 | 0.3-1.5 | 0.46 |
| IL1r4 | 1 | - | 0.46 | 0.359 | 0.334 | 12.3 | 7 | 0.04 | 0.73 | 0.23 | $2.7\cdot10^{-3}$ | 1.5-2.5 | 0.3-0.8 | 0.42 |
| IL2r05 | 8 | 2 (3) | 0.39 | 0.332 | 0.052 | 3.5 | 6.5 | 0.01 | 0.20 | 0.79 | $8.2\cdot10^{-4}$ | 2.1-2.6 | 0.2-1.3 | 0.44 |
| IL2r1 | 5 | (1) | 0.40 | 0.341 | 0.083 | 2.1 | 5 | 0.01 | 0.34 | 0.65 | $1.9\cdot10^{-5}$ | 1.8-2.3 | 0.2-1.0 | 0.51 |
| IL2r4 | 1 | - | 0.59 | 0.439 | 0.104 | 7.1 | 6 | 0.03 | 0.77 | 0.20 | $1.4\cdot10^{-4}$ | 2.3-2.4 | 0.3-1.3 | 0.34 |
| IL3r05 | 9 | 1 (2) | 0.50 | 0.389 | 0.093 | 3.5 | 7 | 0.01 | 0.25 | 0.74 | $1.2\cdot10^{-4}$ | 2.2-2.8 | 0.3-1.3 | 0.39 |
| IL3r1 | 6 | - | 0.57 | 0.363 | 0.081 | 3.1 | 6 | 0.02 | 0.41 | 0.57 | $9.5\cdot10^{-5}$ | 1.9-2.7 | 0.3-1.3 | 0.40 |
| IL3r4 | 1 | 1 (1) | 0.16 | 0.336 | 0.194 | 5.4 | 6 | 0.09 | 0.59 | 0.32 | $9.6\cdot10^{-5}$ | 2.0-2.9 | 0.3-1.2 | 0.81 |

Mercury-like planets are defined as planets obtained after 1 Gyr of orbital evolution having an averaged semi-major axis < 0.5 au and mass > 0.05 $M_\oplus$. NP represents the number of Mercury-like planets identified in individual systems for each simulation type (Sim. ID), and NA gives the total number of Mercury analogs defined by setting the planet's mass as <0.2 (0.25) $M_\oplus$. Also, $m$ is the planet mass, $a$ is the averaged semimajor axis, $e$ is the averaged eccentricity, $i$ is the averaged inclination, nGI is the number of giant impacts (defined by a collision of an embryo/planetesimal that is at least 20% as massive as the target body), and $f_0$, $f_{Emb}$, and $f_{Obj}$ are the mass fractions in the form of the initial embryo, aggregated embryos, and aggregated planetesimals, respectively. WMF is the water mass fraction. All embryos started with an initial WMF = $10^{-5}$ for all planets considered in this table. $a_{0\_H2O}$ indicates the averaged main source region of 90% of the water found in the final planet by the accretion of embryos/planetesimals with an initial WMF > $10^{-5}$ (i.e., objects initially located beyond 1.5 au). Finally, $a_{0\_mass}$ indicates the source region of 90% of the planet's mass, and $f_{MR}$ gives the fraction of the planet's mass sourced by the Mercury region (at 0.2–0.5 au). All $a_{0\_H2O}$ and $a_{0\_mass}$ quantities are given by averaged values, whereas the other quantities (except NP and NA) represent median values for the Mercury-like planets obtained in each simulation type.

**Table 3**. Mercury analogs

| Sim.ID | system # | $m$ (M$_\oplus$) | $a$ (au) | $e$ | $i$ (deg) | nGI | $f_0$ | $f_{Emb}$ | $f_{Obj}$ | WMF | $a_{0\_H2O}$ (au) | $a_{0\_mass}$ (au) | $f_{MR}$ |
|---|---|---|---|---|---|---|---|---|---|---|---|---|---|
| Sr1 | 2 | 0.21 | 0.299 | 0.101 | 4.1 | 2 | 0.084 | 0.188 | 0.728 | $1.8\cdot10^{-5}$ | 1.5-2.1 | 0.2-0.9 | 0.54 |
|  | 3 | 0.23 | 0.301 | 0.208 | 9.6 | 8 | 0.015 | 0.264 | 0.721 | $1.8\cdot10^{-5}$ | 1.5-2.2 | 0.2-1.3 | 0.65 |
| **IL2r05** | **4** | **0.15** | **0.314** | **0.128** | **7.3** | **9** | **0.003** | **0.139** | **0.858** | **$1.7\cdot10^{-3}$** | **2.5-2.6** | **0.2-1.3** | **0.53** |
|  | 2 | 0.18 | 0.308 | 0.107 | 3.2 | 7 | 0.006 | 0.186 | 0.808 | $2.9\cdot10^{-5}$ | 1.6-2.3 | 0.2-1.1 | 0.57 |
|  | **6** | **0.23** | **0.320** | **0.047** | **2.7** | **7** | **0.003** | **0.197** | **0.800** | **$2.0\cdot10^{-5}$** | **1.5-2.2** | **0.2-1.0** | **0.70** |
| **IL2r1** | **2** | **0.23** | **0.292** | **0.130** | **3.3** | **5** | **0.013** | **0.288** | **0.699** | **$1.9\cdot10^{-5}$** | **1.5-2.2** | **0.2-1.1** | **0.70** |
| **IL3r05** | **9** | **0.13** | **0.266** | **0.062** | **4.7** | **7** | **0.010** | **0.141** | **0.849** | **$4.7\cdot10^{-4}$** | **2.2-3.0** | **0.2-1.4** | **0.67** |
|  | 1 | 0.24 | 0.315 | 0.093 | 3.5 | 8 | 0.004 | 0.187 | 0.809 | $3.9\cdot10^{-4}$ | 2.8-2.9 | 0.2-1.1 | 0.49 |
| **IL3r4** | **6** | **0.16** | **0.336** | **0.194** | **5.4** | **6** | **0.086** | **0.590** | **0.324** | **$9.6\cdot10^{-5}$** | **2.0-2.9** | **0.3-1.2** | **0.81** |

Mercury analogs are defined as planets obtained after 1 Gyr of orbital evolution having an averaged semi-major axis < 0.5 au and 0.05 < mass < 0.25 M$_\oplus$. All variables as defined in the caption of Table 2; however, unlike in that table, the values above are given for nine individual Mercury analog planets. Also, the planets obtained in a particular simulation come from distinct runs of that simulation type (Sim.ID), as indicated by the system number # shown in Fig. 3. Finally, the best system analogs are highlighted in bold (see Section 3.3 for details).

**Table 4.** Venus analogs obtained in the systems with Mercury analogs

| Sim.ID | system # | $m$ (M$_\oplus$) | $a$ (au) | $e$ | $i$ (deg) | nGI | $f_0$ | $f_{\text{Emb}}$ | $f_{\text{Obj}}$ | WMF | $a_{0\_\text{H2O}}$ (au) | $a_{0\_\text{mass}}$ (au) | $f_{\text{MR}}$ |
|---|---|---|---|---|---|---|---|---|---|---|---|---|---|
| Sr1 | 2 | 0.82 | 0.461 | 0.062 | 2.0 | 7 | 0.016 | 0.424 | 0.560 | $5.8 \cdot 10^{-5}$ | 1.9-2.7 | 0.3-1.1 | 0.34 |
| | 3 | 0.90 | 0.455 | 0.080 | 3.6 | 8 | 0.011 | 0.443 | 0.546 | $5.2 \cdot 10^{-5}$ | 1.8-2.6 | 0.3-1.0 | 0.32 |
| **IL2r05** | **4** | **1.44** | **0.519** | **0.057** | **2.0** | **9** | 0.003 | 0.307 | 0.690 | **$1.1 \cdot 10^{-3}$** | **2.4-3.4** | **0.3-1.3** | **0.21** |
| | 2* | 0.72 | 0.446 | 0.031 | 2.0 | 6 | 0.013 | 0.237 | 0.750 | $1.4 \cdot 10^{-5}$ | 1.5-2.1 | 0.3-1.2 | 0.43 |
| | **6** | **1.19** | **0.537** | **0.028** | **1.8** | **7** | 0.008 | 0.318 | 0.674 | **$4.0 \cdot 10^{-4}$** | **2.2-2.9** | **0.3-1.7** | **0.21** |
| **IL2r1** | **2** | **1.44** | **0.541** | **0.033** | **2.0** | **7** | 0.013 | 0.488 | 0.499 | **$6.0 \cdot 10^{-4}$** | **2.0-3.1** | **0.3-1.3** | **0.19** |
| **IL3r05** | **9** | **1.00** | **0.547** | **0.020** | **2.0** | **7** | 0.005 | 0.310 | 0.685 | **$6.6 \cdot 10^{-5}$** | **1.8-2.6** | **0.3-1.4** | **0.24** |
| | 1 | 1.06 | 0.488 | 0.035 | 1.9 | 9 | 0.004 | 0.298 | 0.698 | $1.1 \cdot 10^{-4}$ | 1.8-2.9 | 0.3-1.4 | 0.29 |
| **IL3r4** | **6** | **1.18** | **0.502** | **0.067** | **1.9** | **7** | 0.033 | 0.770 | 0.197 | **$9.4 \cdot 10^{-5}$** | **2.0-2.1** | **0.3-1.4** | **0.19** |

Venus analogs are defined as planets obtained after 1 Gyr of orbital evolution with a semi-major axis larger than that of the Mercury analog and $0.5 < \text{mass} < 1.5$ M$_\oplus$. All variables are as defined in the caption of Table 2; however, unlike in that table, the values above are given for nine individual Venus-analog planets. Also, the planets obtained in a particular simulation come from distinct runs of that simulation type (Sim.ID), as indicated by the system number # shown in Fig. 3. Finally, the best system analogs are highlighted in bold (see Section 3.3 for details).

* PS: System #2 of IL2r05 contains a second Venus-like planet.